\documentclass[a4paper,11pt]{article}
\usepackage{pos}
\usepackage{subcaption}

\usepackage{xspace}

\newcommand{\aunit}[1]{\ensuremath{\text{\,#1}}} 
\newcommand{\gevcc}{\ensuremath{\aunit{Ge\kern -0.1em V\!/}c^2}\xspace}
\newcommand{\gevc}{\ensuremath{\aunit{Ge\kern -0.1em V\!/}c}\xspace}
\def\sqsnn {\ensuremath{\protect\sqrt{s_{\scriptscriptstyle\text{NN}}}}\xspace}
\def\nb {\aunit{nb}\xspace}
\def\invnb {\ensuremath{\nb^{-1}}\xspace}
\newcommand{\tev}{\aunit{Te\kern -0.1em V}\xspace}

\def\Zcj{\ensuremath{\mathcal{R}^c\kern-0.4em{\raisebox{-0.2em}{$\scriptstyle j$}}}\xspace}

\def\to                 {\ensuremath{\rightarrow}\xspace}
\def\Pmu         {\ensuremath{\mu}\xspace}
\def\mumu       {{\ensuremath{\Pmu^+\Pmu^-}}\xspace}

\def\PZ  {\ensuremath{\mathrm{Z}}\xspace} 
\def\Z      {{\ensuremath{\PZ}}\xspace}

\def\gevcsquare    {\ensuremath{{\,(\mathrm{Ge\kern -0.1em V\!/}c)^2}}\xspace}
\def\mygevc        {\ensuremath{{\mathrm{Ge\kern -0.1em V\!/}c}}\xspace}
\def\unitgev   {\ensuremath{\mathrm{[Ge\kern -0.1em V]}}\xspace}

\def\pPb   {\ensuremath{p\xspace\mathrm{Pb}}\xspace}

\def\PbPb   {\ensuremath{\mathrm{Pb\xspace Pb}}\xspace}

\def\rfb   {\ensuremath{R_\mathrm{FB}}\xspace}
\def\rpa   {\ensuremath{R_{\pPb}}\xspace}

\def\Zmumu   {\ensuremath{\Z\to\mumu}\xspace}

\def\phistar {\ensuremath{\phi^{*}}\xspace}

\def\zrapstar {\ensuremath{y^{*}_{\Z}}\xspace}
\def\ZpT     {\ensuremath{p_{\mathrm{T}}^{\Z}}\xspace}

\def\powheg     {\mbox{\textsc{PowhegBox}}\xspace}

\def\coh{\mathrm{coh}}
\def\logpt2 {\ensuremath{\ln(\pt^{*2})}\xspace} 

\def\gevcsquare    {\ensuremath{{\,(\mathrm{Ge\kern -0.1em V\!/}c)^2}}\xspace}
\def\mygevc        {\ensuremath{{\mathrm{Ge\kern -0.1em V\!/}c}}\xspace}
\def\unitmevc   {\ensuremath{\mathrm{[Me\kern -0.1em V\!/}c]}\xspace}
\def\unitgevc   {\ensuremath{\mathrm{[Ge\kern -0.1em V\!/}c]}\xspace}
\def\unitmbarngevc{\ensuremath{[\mathrm{mb/(Ge\kern -0.1em V\!/}c)]}\xspace}

\def\PJ      {\ensuremath{\mathrm{J}}\xspace}
\def\Ppsi        {\ensuremath{\psi}\xspace} 

\def\jpsi     {{\ensuremath{{\PJ\mskip -3mu/\mskip -2mu\Ppsi\mskip 2mu}}}\xspace}
\def\psitwos  {{\ensuremath{\Ppsi{(2S)}}}\xspace}

\def\pt         {\ensuremath{p_{\mathrm{T}}}\xspace}

\newcommand{\nospaceunit}[1]{\ensuremath{\text{#1}}}       

\def\mub{\ensuremath{\,\upmu\nospaceunit{b}}\xspace}
\def\mbarn{\aunit{mb}\xspace}

\newcommand{\etal}{\mbox{\itshape et al.}\xspace}

\title{LHCb measurements of Quarkonia Production in Ultraperipheral PbPb collisions and Z production in $p$Pb collisions}

\author*[a,b]{Hengne Li} 

\affiliation[a]{on behalf of the LHCb collaboration.}
\affiliation[b]{Guangdong Provincial Key Laboratory of Nuclear Science, Guangdong-Hong Kong Joint Laboratory of Quantum Matter, Institute of Quantum Matter, South China Normal University, Guangzhou, China.}

\emailAdd{hengne.li@m.scnu.edu.cn}

\abstract{Measurements of quarkonia production in ultra-peripheral heavy-ion collisions are of important value to study photon-photon and photon-nucleus interactions, the partonic structure of nuclei, and mechanisms of vector-meson production. LHCb has studied both coherent 
\jpsi
and
\psitwos
mesons in ultra-peripheral collisions using PbPb data at forward rapidity with the highest precision currently accessible. In addition, measurements of 
\Z
production in 
\pPb
collisions provide new constraints on the partonic structure of nucleons bound inside nuclei. Here will present these measurements of quarkonia and 
\Z 
production, along with comparisons with the latest theoretical models.}

\FullConference{%
  41st International Conference on High Energy physics - ICHEP2022\\
  6-13 July, 2022\\
  Bologna, Italy
}

\begin{document}
\maketitle

\section{Introduction}

The LHCb detector~\cite{LHCb-DP-2008-001, LHCb-DP-2014-002} 
is a fully instrumented single-arm spectrometer in the forward region 
covering a pseudorapidity acceptance of
$2 < \eta < 5$, 
providing precise vertex reconstruction and excellent tracking momentum resolution down to a very low 
transverse momentum (\pt). 
The heavy ion program at LHCb includes both proton-lead (\pPb) and lead-lead (\PbPb) collisions.
Forward acceptance at LHCb allows to study parton distribution functions in protons and nuclei in both small ($x < 10^{-3}$) and large ($10^{-1}<x<1$) Bjorken-$x$ regions.
The coherent (exclusive) photo-production of charmonium in \PbPb ultra-peripheral collisions (UPC) provides an ideal laboratory to 
probe the nuclear
gluon
distribution functions at a momentum transfer of 
$Q^2\approx m^2/4$, where $m$ is the mass of the meson.
The electroweak \Z-boson production and its leptonic decay products once produced do not participate in hadronic interactions, are ideal probes of the nuclear modifications in the non-perturbative initial-state.
Two recent results are presented in this article,  the measurement of coherent \jpsi and
\psitwos production in \PbPb UPC~\cite{LHCb:2022ahs} and the measurement of the \Z production in \pPb collisions~\cite{LHCb:2022kph}.

\section{Charmonium production in ultra-peripheral \PbPb collisions}

The coherent $\jpsi$ and $\psitwos$ production are measured in \PbPb UPC at a nucleon-nucleon centre-of-mass energy of $5.02\tev$ using a data sample corresponding to an integrated luminosity of $228\pm10\mub^{-1}$, collected by the LHCb experiment in 2018~\cite{LHCb:2022ahs}. 

The integrated cross-sections of 
coherent $\jpsi$ and $\psitwos$ 
production in \PbPb collisions are measured 
in the rapidity region $2.0 < y^* < 4.5$ as
\begin{align*}
\sigma^\coh_{\jpsi} &= 5.965 \pm 0.059 \pm 0.232 \pm 0.262 \mbarn\,, \\
\sigma^\coh_{\psitwos} &= 0.923 \pm 0.086 \pm 0.028 \pm 0.040 \mbarn\,, 
\end{align*}
where the first listed uncertainty
is statistical, 
the second is systematic
and the third is due to
the luminosity determination.
The ratio of the coherent 
\psitwos to \jpsi production cross-sections is measured to be
\begin{align*}
\sigma^\coh_{\psitwos}/\sigma^\coh_{\jpsi} &= 0.155 \pm 0.014 \pm 0.003\,,
\end{align*}
where the first uncertainty is 
statistical and the second is systematic. 
The luminosity uncertainty cancels 
in the ratio measurement.

\begin{figure}[ht]
    \centering
    \hfil
    \begin{minipage}[t]{0.33\linewidth}
        \centering
        \includegraphics[width=\linewidth]{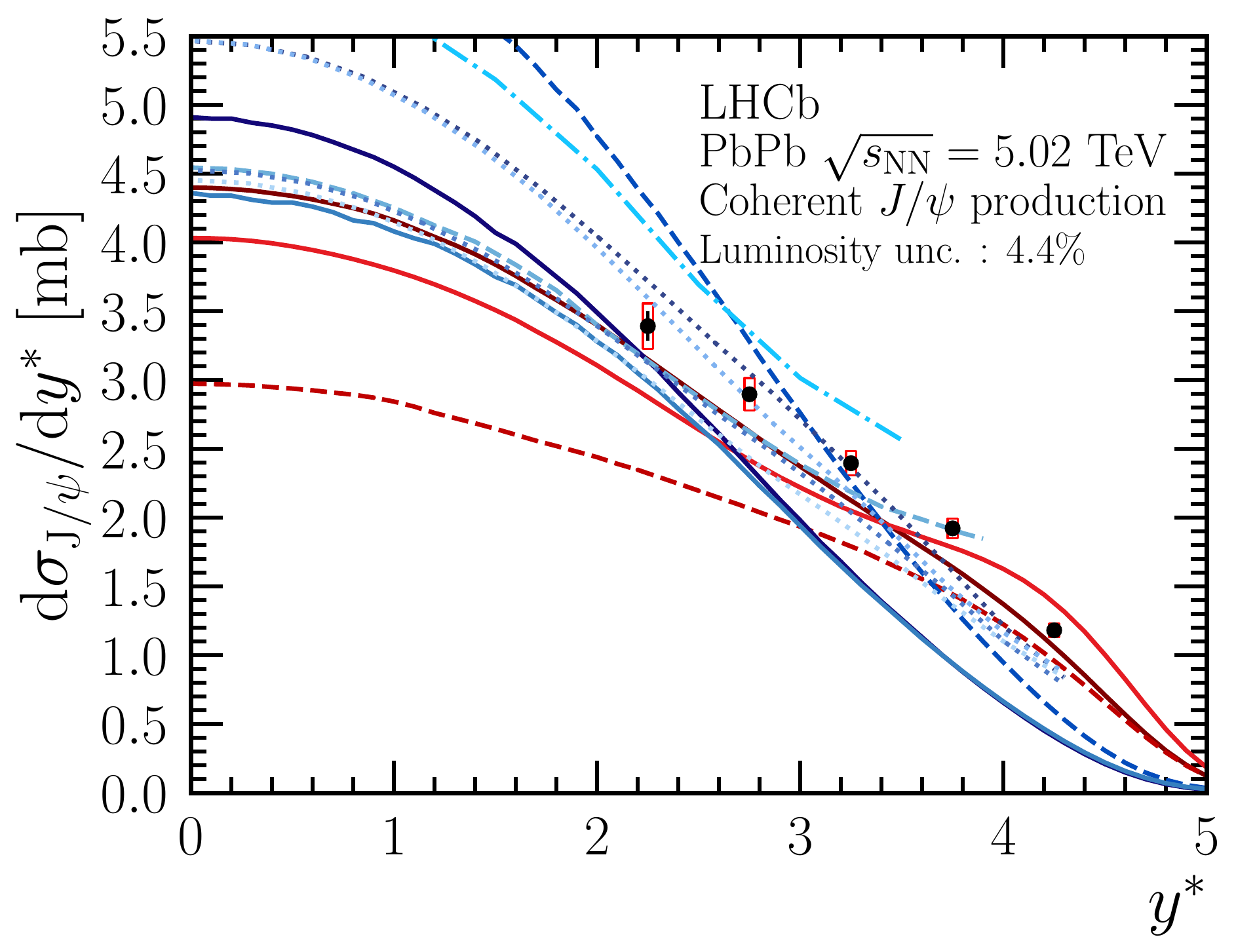}
    \end{minipage}
    \begin{minipage}[t]{0.429\linewidth}
        \centering
        \includegraphics[width=\linewidth]{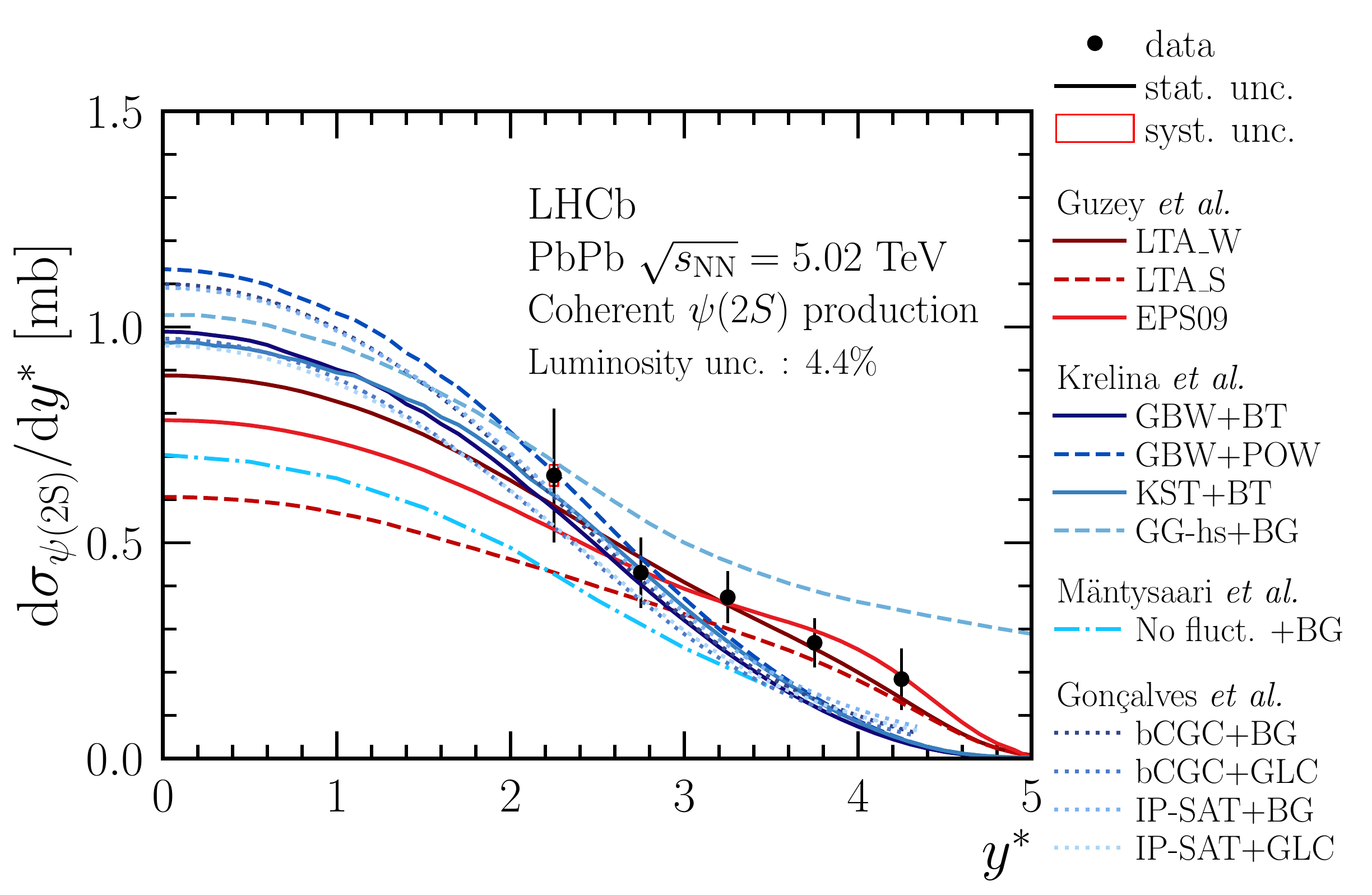}
     \end{minipage}
     \hfil
     \vspace{-0.3cm}
    \caption{Differential cross-section as a function $y^*$ for coherent (left) $\jpsi$ and (right) \psitwos production, compared to theoretical predictions. These models are grouped as (red lines) perturbative-QCD calculations and (blue lines) colour-glass-condensate models.}
    \label{fig:theo_y}
\end{figure}

\begin{figure}[htbp]
    \centering
    \hfil
    \begin{minipage}[t]{0.33\linewidth}
        \centering
        \includegraphics[width=\linewidth]{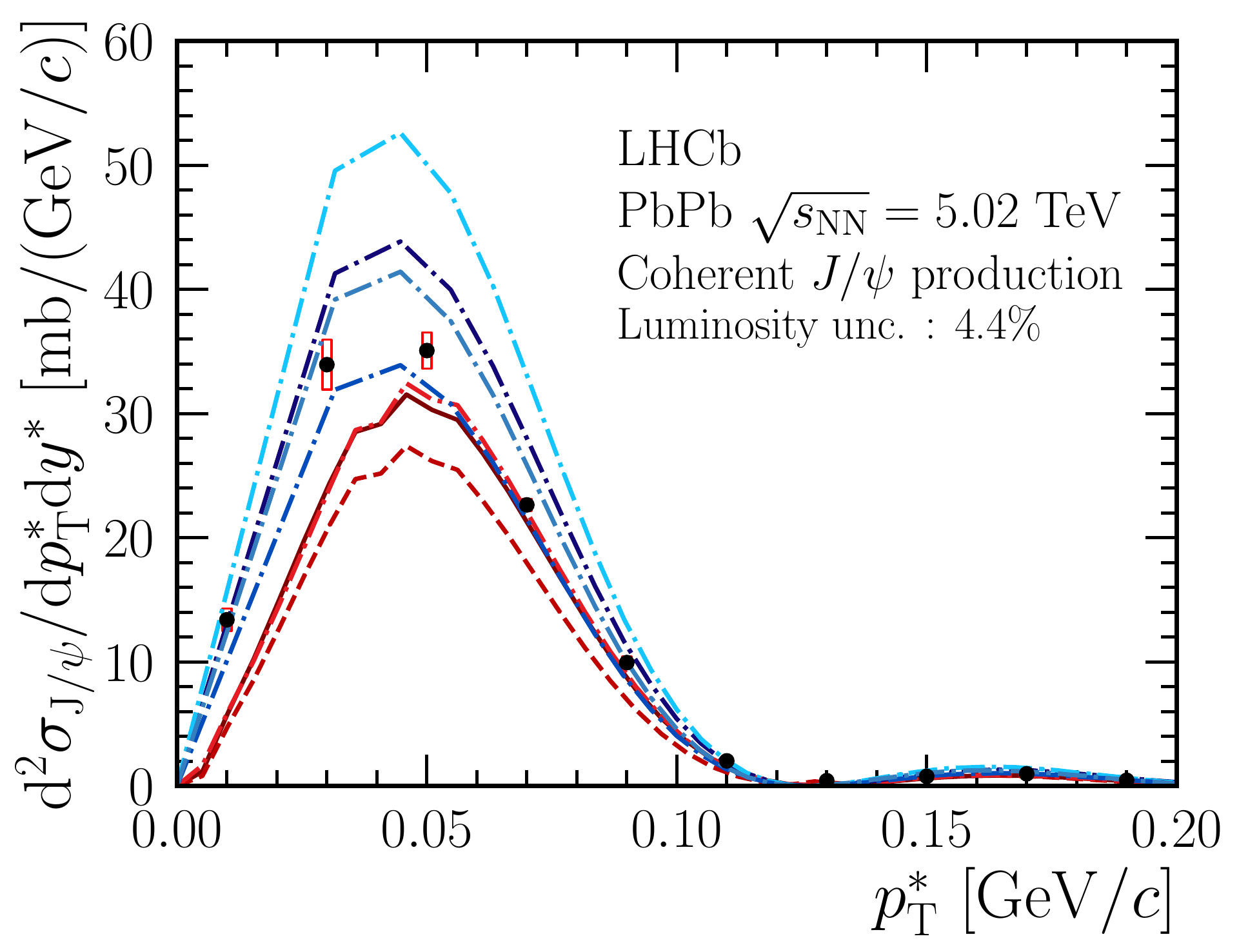}
    \end{minipage}
    \begin{minipage}[t]{0.429\linewidth}
        \centering
        \includegraphics[width=\linewidth]{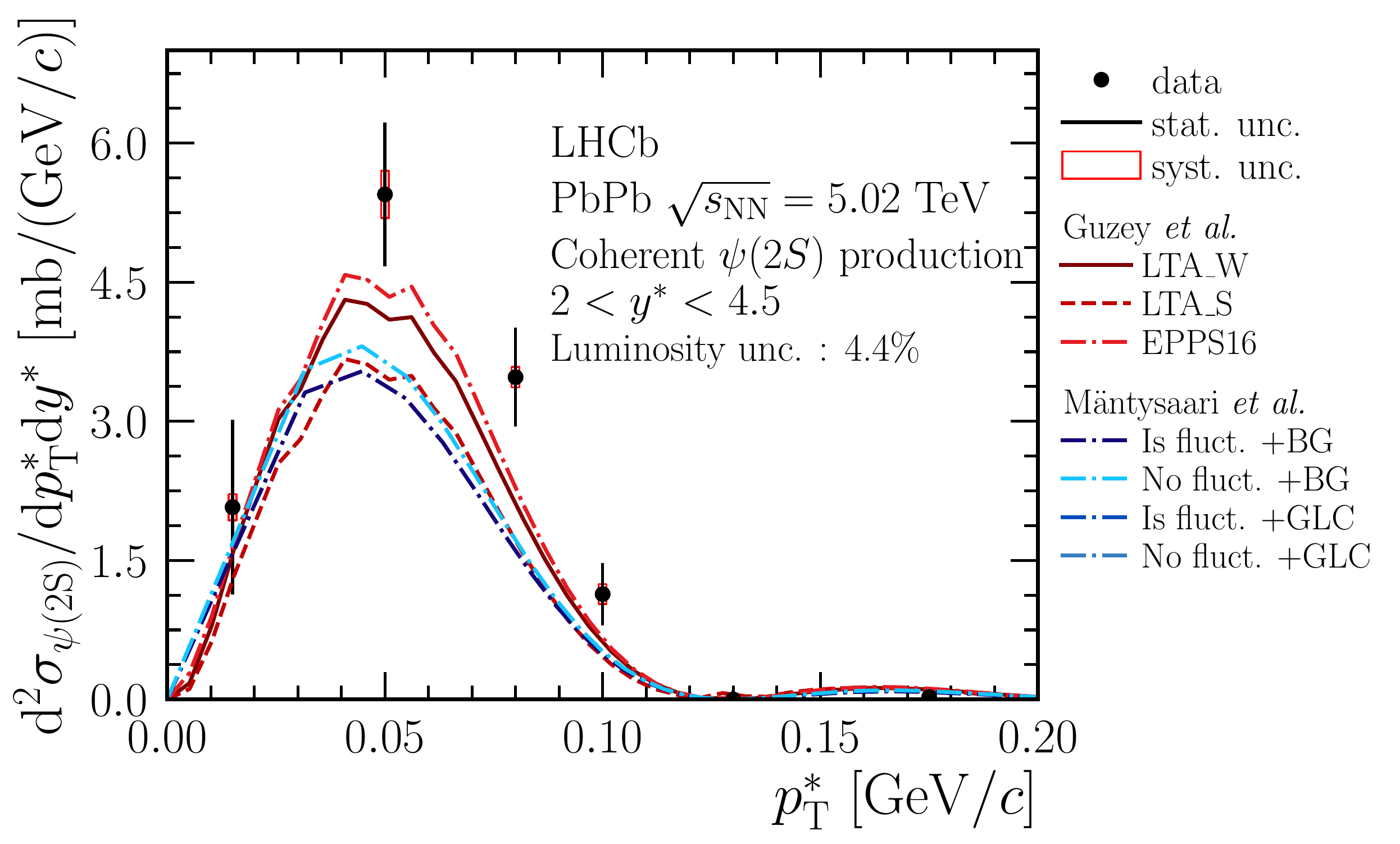}
    \end{minipage}
    \hfil
    \vspace{-0.3cm}
    \caption{Differential cross-section as a function of $\pt^*$ within the rapidity range $2<y^*<4.5$ for coherent (left) $\jpsi$ and (right) \psitwos production compared to theoretical predictions.}
    \label{fig:theo_pt}
\end{figure}

\begin{figure}[htbp]
\begin{center}
\includegraphics[width=0.45\linewidth]{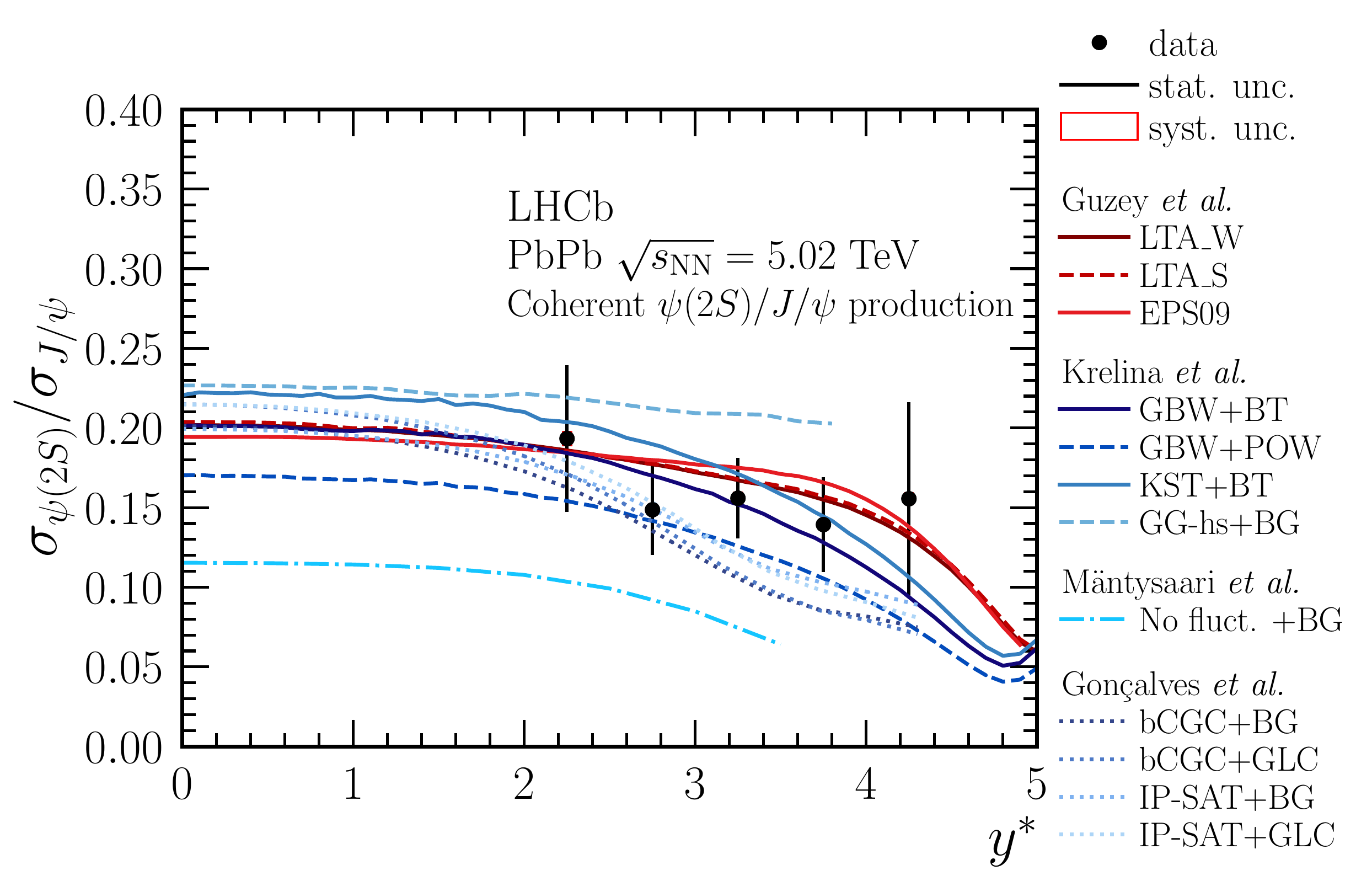}
\end{center}
\vspace{-0.7cm}
\caption{Differential cross-section ratio of \psitwos to \jpsi as a function of $y^*$, compared to theoretical predictions. These models are separated into (red lines) perturbative QCD calculations and (blue lines) colour-glass-condensate models.}
\label{fig:y_theo_ratio}
\end{figure}

The measured differential 
cross-sections as a function of $y^*$
and $\pt^*$ for coherent \jpsi and \psitwos 
are shown in 
Figs.~\ref{fig:theo_y}
and \ref{fig:theo_pt}, respectively.
The cross-section ratio of coherent 
\psitwos to \jpsi production as 
a function of rapidity is measured for the first time and shown in
Fig.~\ref{fig:y_theo_ratio}. 
These results are compared to several theoretical
predictions
which can be grouped into models based on
perturbative-QCD (pQCD)~\cite{Guzey_2016,2017access} 
and colour-glass-condensate (CGC) 
calculations~\cite{PhysRevC.84.011902,2018,
Kopeliovich:2020has,PhysRevD.96.094027,
Gon_alves_2005,20171,Mantysaari:2017dwh,
2014arXiv1406.2877L}~\footnote{New theoretical developments after the Conference talk are not yet included in the figures, but they are well noticed and acknowledged.}.


The models provided by 
Guzey \etal~\cite{Guzey_2016,2017access} 
are based on pQCD calculations under the leading-logarithm approximation. 
They are compatible with the data, 
with excellent agreement
at high rapidity and a slight trend of
underestimation at low
rapidity for both \jpsi and \psitwos,
as shown in Fig.~\ref{fig:theo_y}. 
The underestimation at low rapidity 
results in an overall lower prediction
in the differential cross-section as 
a function of $\pt^*$,
shown in Fig.~\ref{fig:theo_pt}. 
However, 
excellent agreement between the prediction and data for the 
\psitwos to \jpsi ratio measurement
can be observed in Fig.~\ref{fig:y_theo_ratio}.

The models by 
Krelina \etal~\cite{Kopeliovich:2020has,2018},
M\"antysaari \etal~\cite{20171,2014arXiv1406.2877L,Mantysaari:2017dwh},
and 
Gon\c{c}alves \etal~\cite{PhysRevD.96.094027,Gon_alves_2005} 
can be considered as variations of the 
colour-dipole model
based on CGC theory. 
These models are compatible with the data,
with large variations between different models.
Better agreement can be seen at low rapidity than at high rapidity.
A relatively larger underestimation
at high rapidity and 
as a function of $\pt^*$
can be seen for the \psitwos meson in 
Figs.~\ref{fig:theo_y} and
\ref{fig:theo_pt}.
This can be understood as causing by
higher theoretical uncertainty,
since the wave function of the \psitwos meson
has a more complicated structure
than that of the \jpsi~\cite{20171,2014arXiv1406.2877L,Kowalski:2006hc}.

\section{Production of \Z-boson in \pPb collisions}

The \Zmumu production is measured~\cite{LHCb:2022kph} using \pPb collision
dataset collected at $\sqsnn=8.16\tev$ in 2016 by the LHCb detector corresponding to an integrated luminosity of $12.2\pm0.3\invnb$ for forward collisions
and $18.6\pm0.5\invnb$ for backward collisions.
The differential cross-section,
the forward-backward ratio (\rfb) of the production cross-sections and the nuclear modification factors (\rpa) are measured for the first time as a function of the rapidity of the \Z boson in the centre-of-mass frame (\zrapstar),
the transverse momentum (\ZpT) and an angular variable \phistar~\cite{Vesterinen:2008hx}.

\begin{figure}[htbp]
\begin{center}
\begin{subfigure}[b]{0.32\textwidth}
\centering
\includegraphics[width=\textwidth]{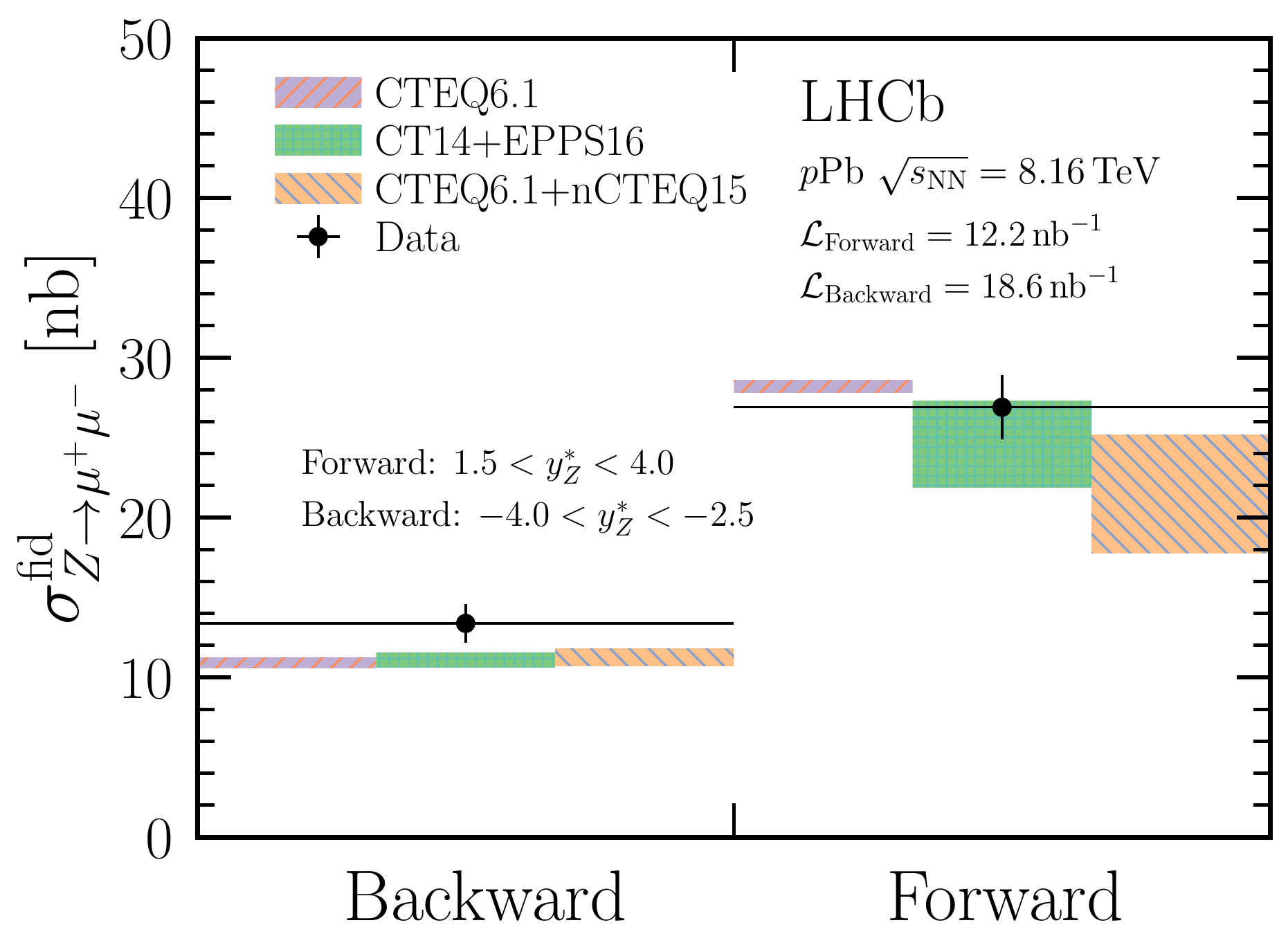}
\put(-30,20){(a)}
\end{subfigure}
\begin{subfigure}[b]{0.32\textwidth}
\centering
\includegraphics[width=\textwidth]{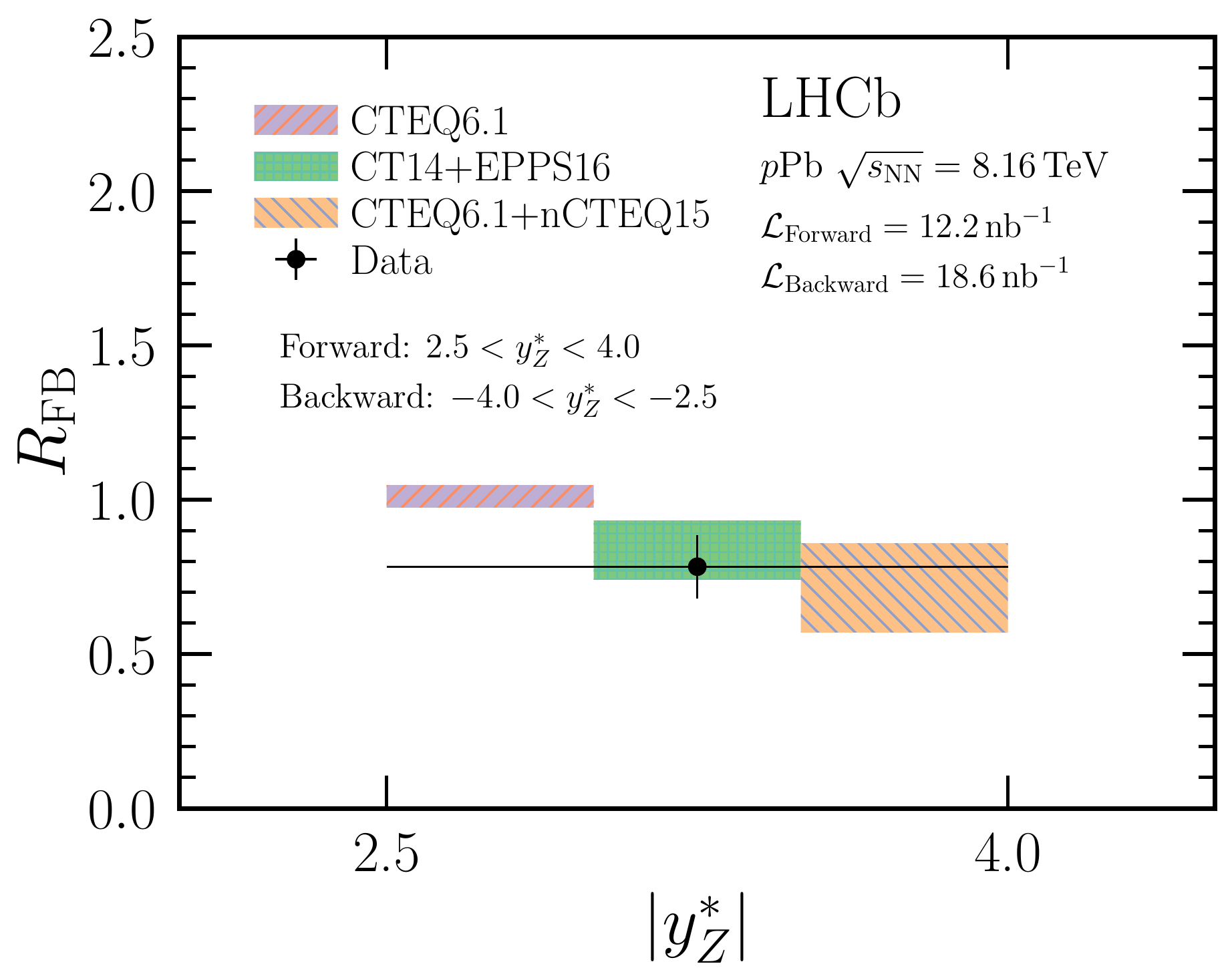}
\put(-30,25){(b)}
\vspace*{-.21cm}
\end{subfigure}
\begin{subfigure}[b]{0.32\textwidth}
\centering
\includegraphics[width=\textwidth]{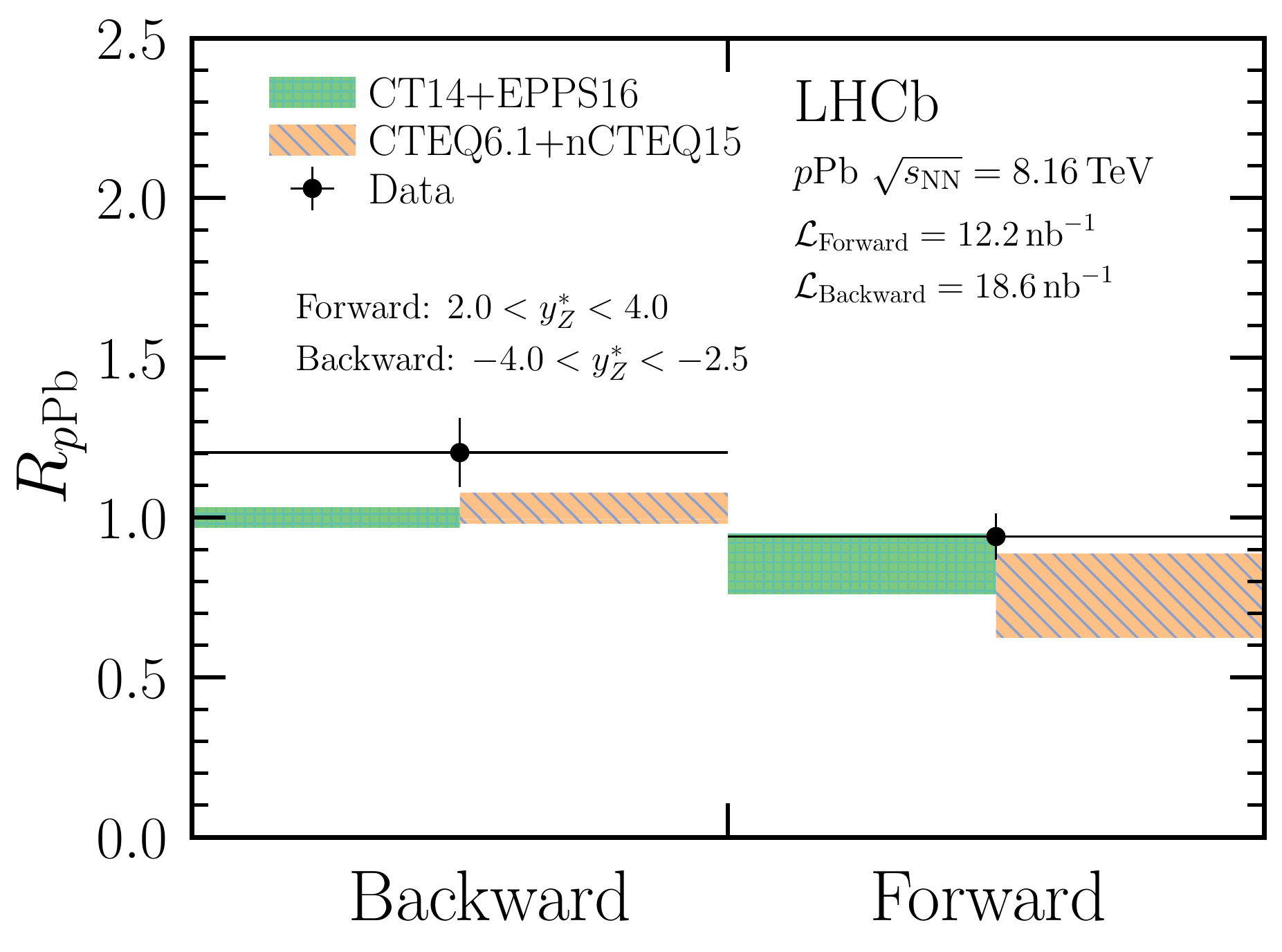}
\put(-30,20){(c)}
\end{subfigure}
\end{center}
\vspace*{-0.5cm}
\caption{
The measured inclusive (a) \Zmumu production fiducial cross-section, (b) forward-backward ratio (\rfb), and (c) nuclear modification factors (\rpa),  
compared to theoretical predictions.
}
\vspace*{-0.5cm}
\label{fig:zmmtotal}
\end{figure}

The measured inclusive fiducial cross-section, forward-backward ratio (\rfb), and nuclear modification factors (\rpa) are shown in Fig.~\ref{fig:zmmtotal}, together with
the comparisons to the \powheg prediction using
CTEQ6.1, EPPS16 and nCTEQ15 (n)PDF sets,
for forward and backward collisions, respectively.
For forward collisions,
the measured cross-section shown in Fig.~\ref{fig:zmmtotal}(a) appears to have a good agreement
with the \powheg calculations, with a smaller uncertainty
for the two intervals of $2.0<\zrapstar<3.0$ compared
to the theoretical calculations,
which can be used to further constrain the
nPDFs.
For backward collisions,
the uncertainty of the measurement is larger than
that of the \powheg calculation, and the measured central value is higher than the prediction especially
for the $-3.5 <\zrapstar <-3.0$ interval by about 2$\sigma$.
However, the measurement and calculation are compatible.
The measured value of \rfb shown in Fig.~\ref{fig:zmmtotal}(b) is below unity,
which is a reflection of the suppression
due to, {\it e.g.}, nuclear shadowing at small
Bjorken-$x$, together with an average
enhancement at large Bjorken-$x$.
The data is in agreement
with the EPPS16 and nCTEQ15 predictions.
The uncertainty of the measurement is
smaller than the theoretical uncertainties
using EPPS16 and nCTEQ15 nPDFs,
showing a constraining power on the
nPDFs.
The measured overall \rpa values are shown in Fig.~\ref{fig:zmmtotal}(c) and are 
compared to the \powheg predictions using the
EPPS16 and nCTEQ15 nPDF sets.
The overall \rpa results show good compatibility
between measurements and theoretical predictions.
The backward rapidity result shows
larger uncertainty compared to that of the
nPDF sets.
The measured central value is consistent with the
prediction at a 2$\sigma$ level.
The forward rapidity result
gives a higher precision than the
EPPS16 and nCTEQ15 nPDF sets,
and the central value is larger than
the prediction,
which shows a constraining power
on the current nPDF sets.

\begin{figure}[htbp]
\begin{center}
\begin{subfigure}[b]{0.305\textwidth}
\centering
\includegraphics[width=\textwidth]{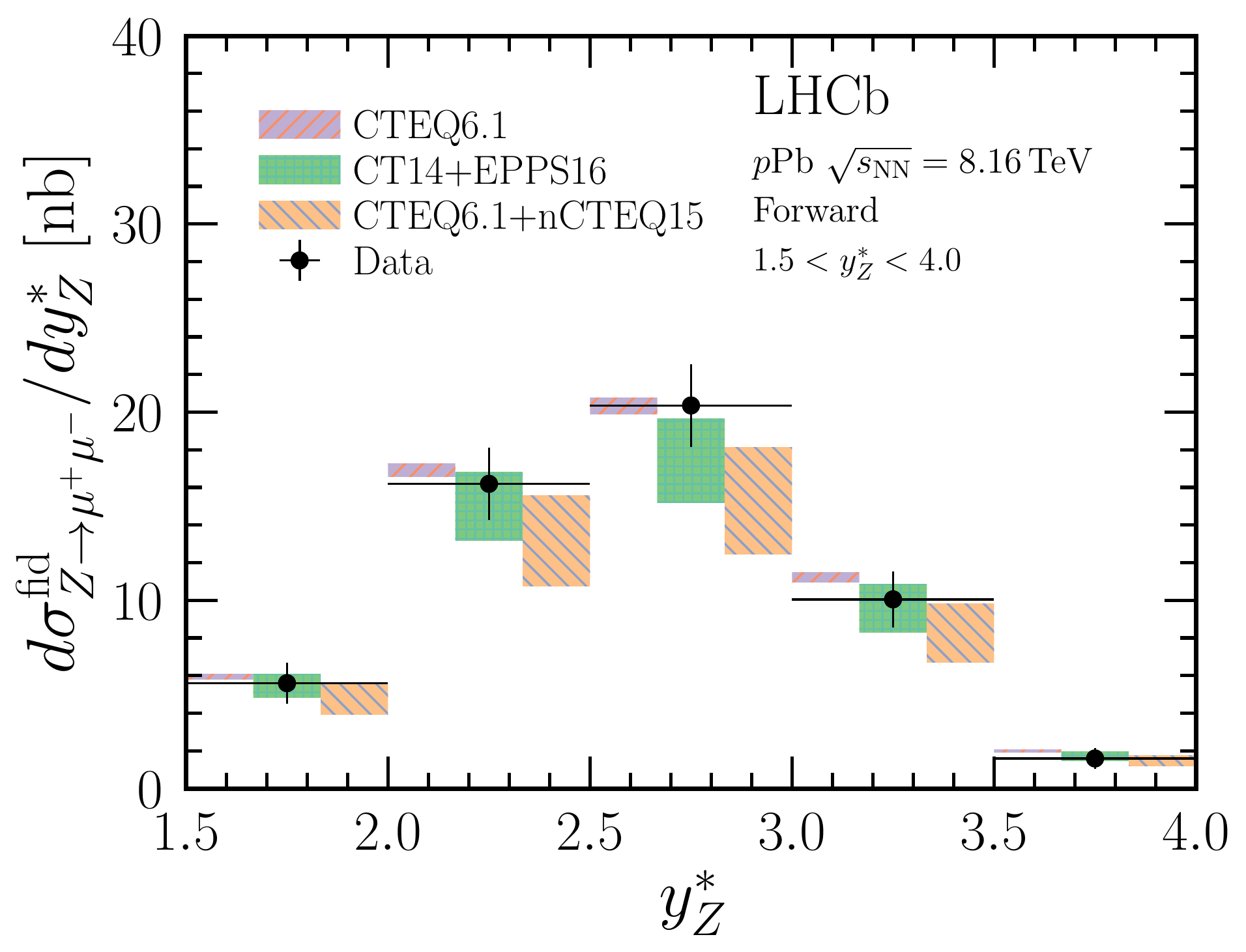}
\put(-25,50){(a)}
\end{subfigure}
\begin{subfigure}[b]{0.305\textwidth}
\centering
\includegraphics[width=\textwidth]{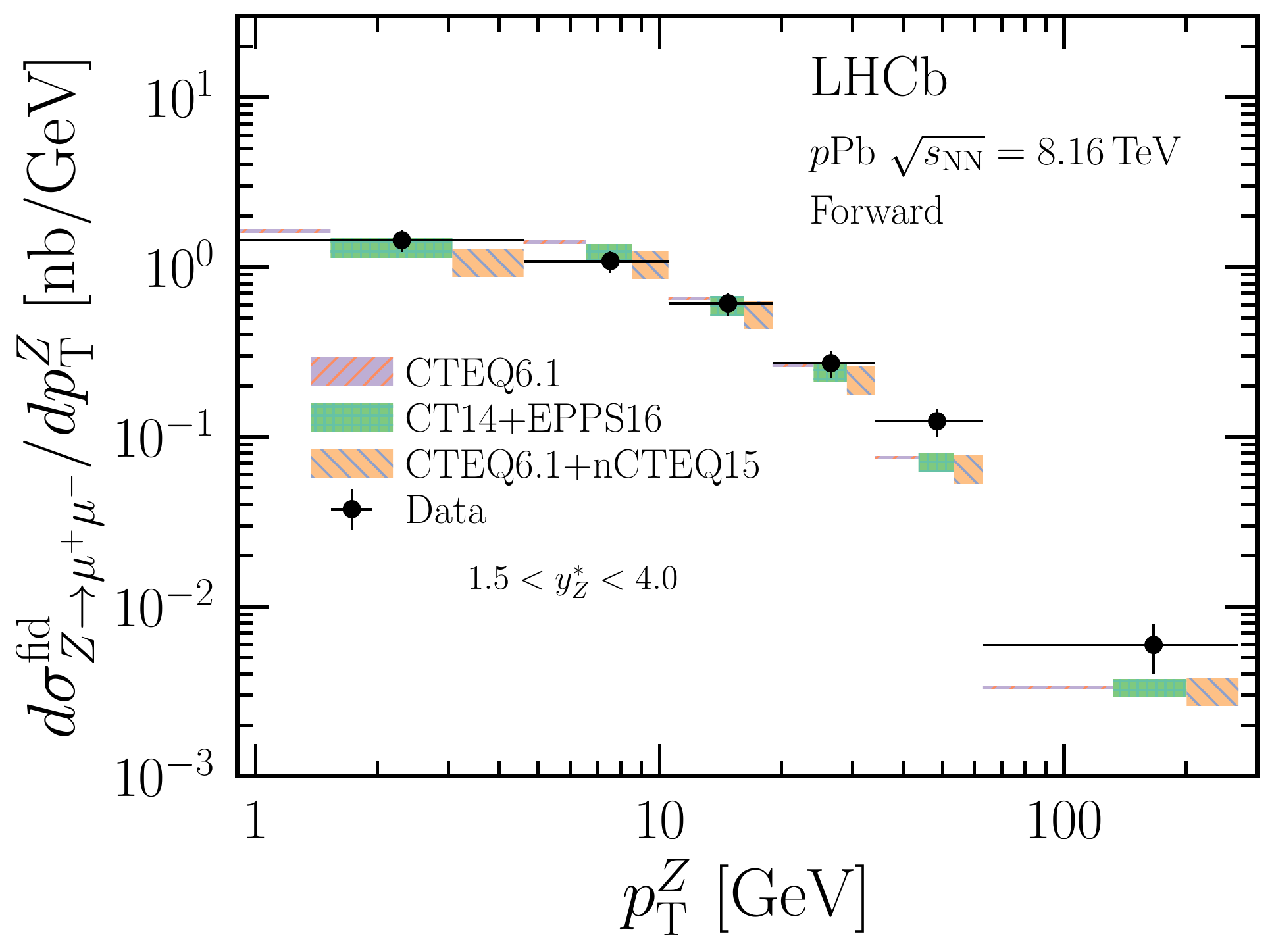}
\put(-25,50){(b)}
\vspace*{-.05cm}
\end{subfigure}
\begin{subfigure}[b]{0.32\textwidth}
\centering
\includegraphics[width=\textwidth]{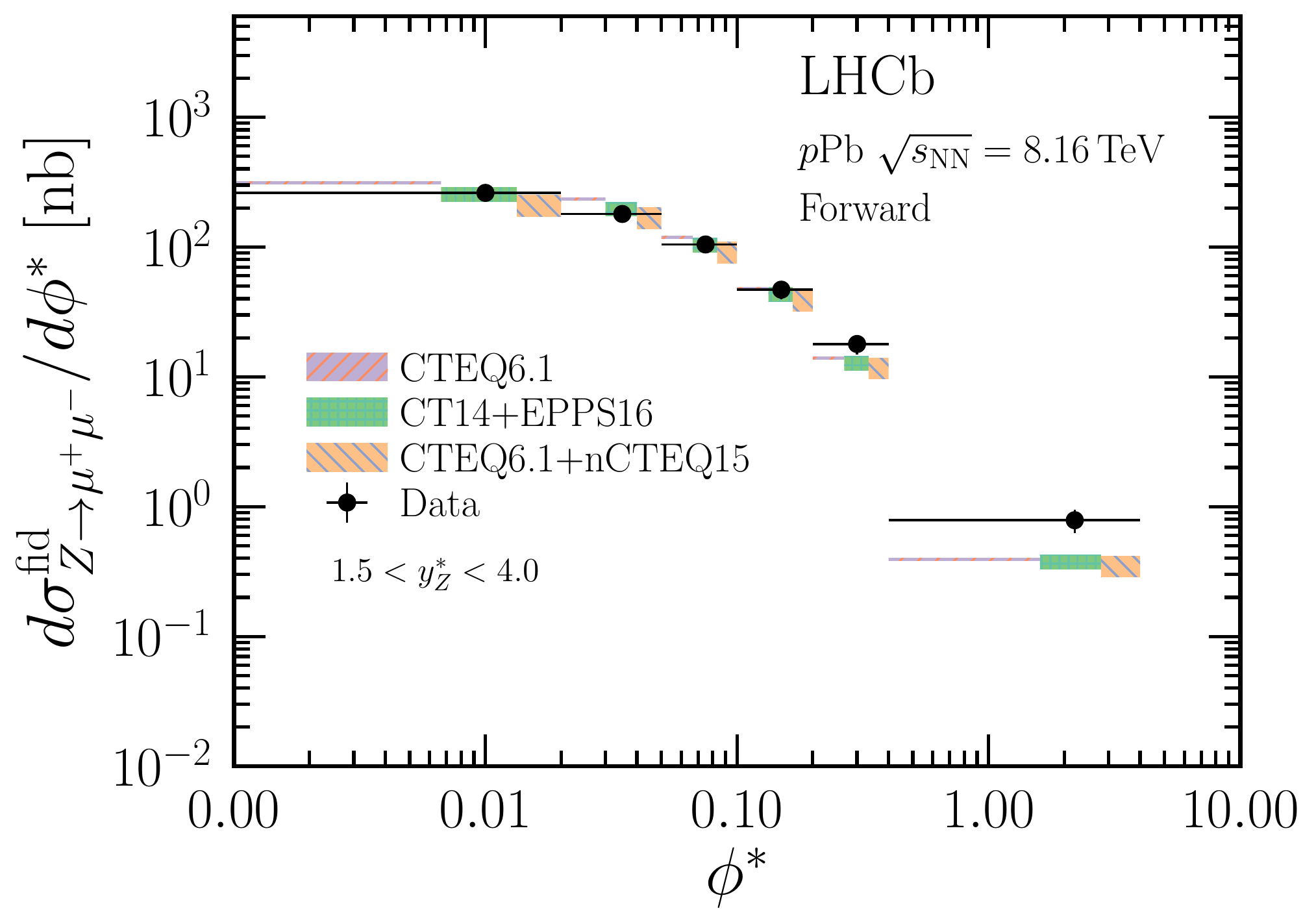}
\put(-25,50){(c)}
\end{subfigure}
\begin{subfigure}[b]{0.31\textwidth}
\centering
\includegraphics[width=\textwidth]{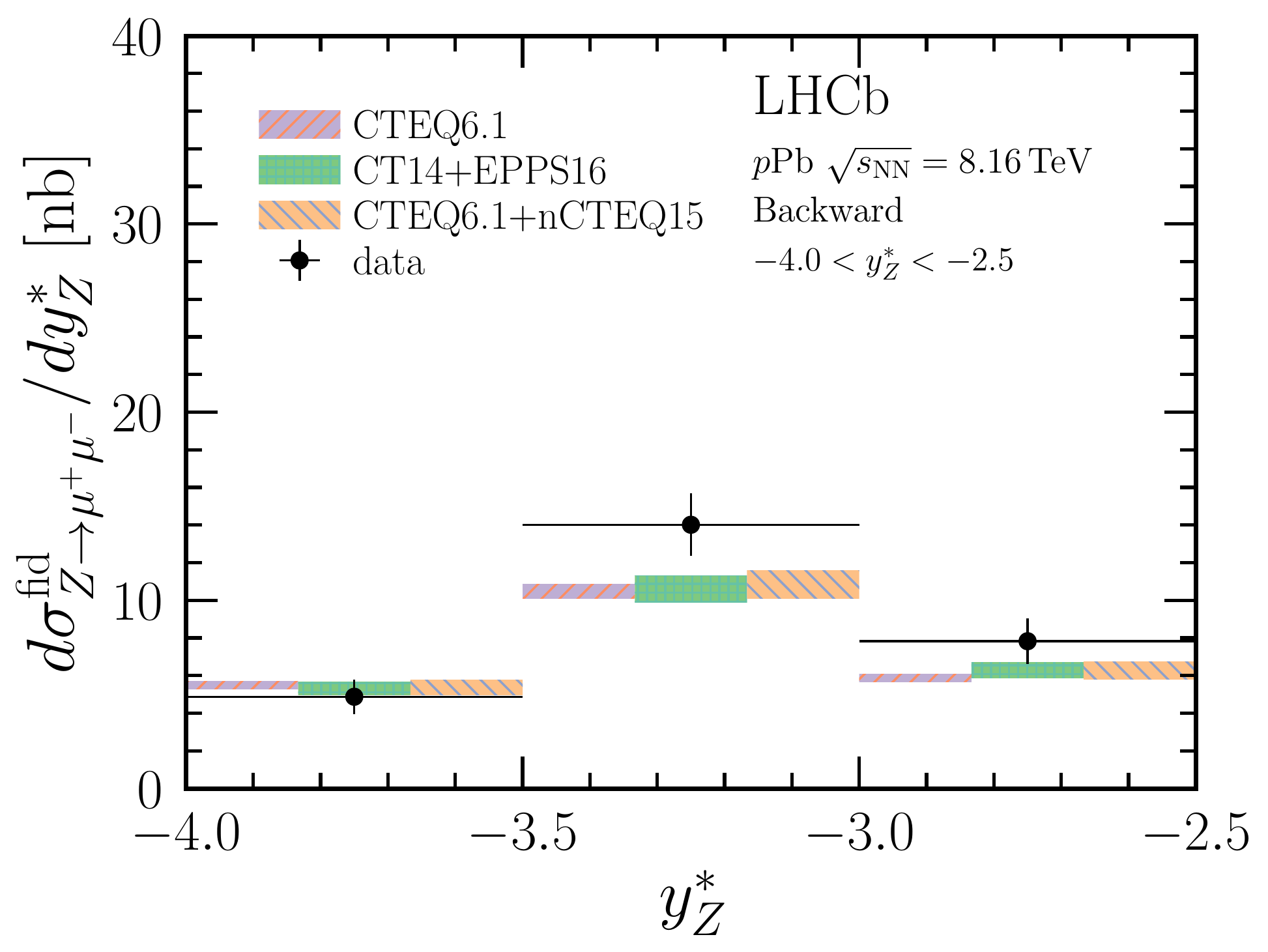}
\put(-25,50){(d)}
\end{subfigure}
\begin{subfigure}[b]{0.305\textwidth}
\centering
\includegraphics[width=\textwidth]{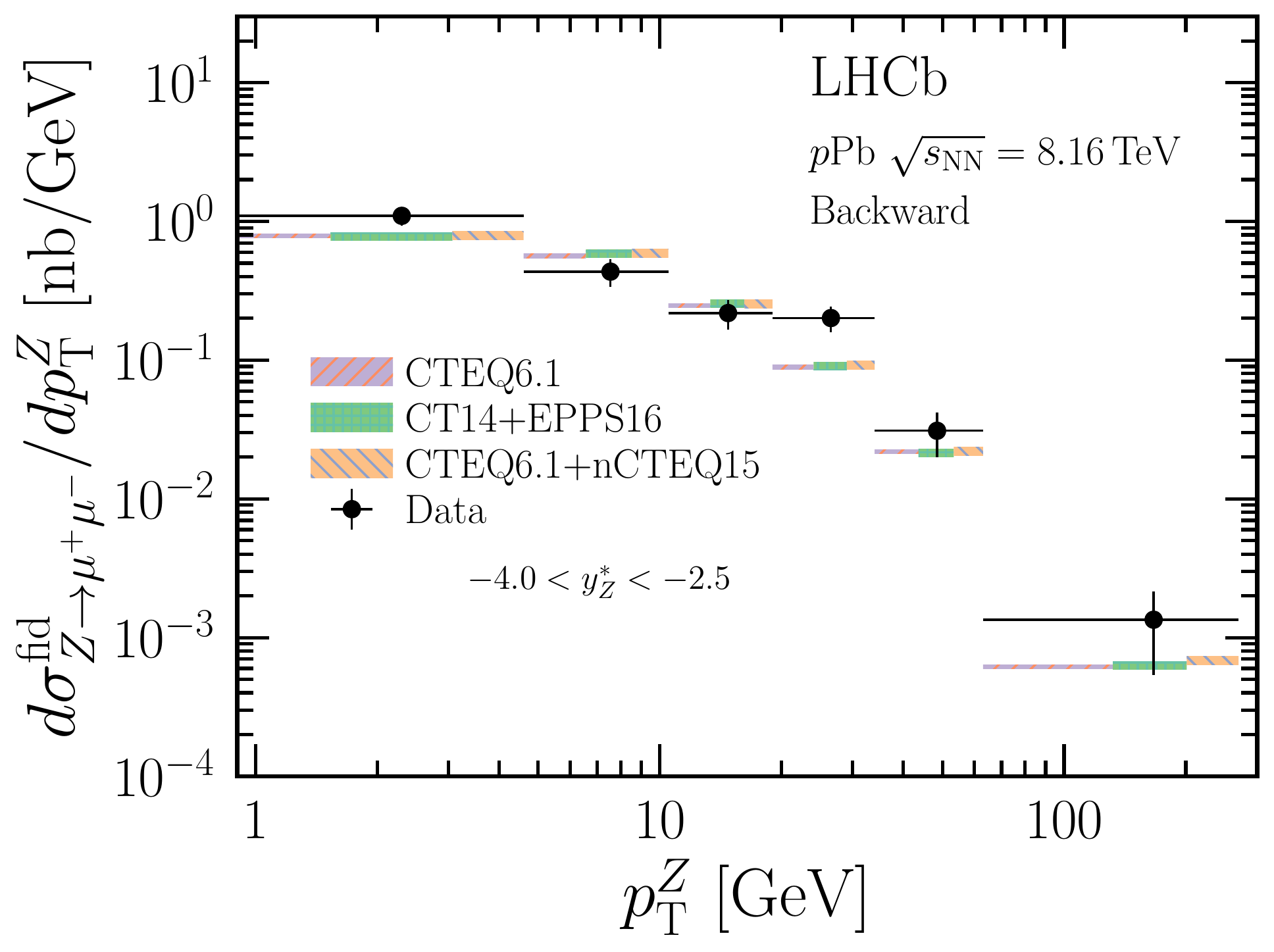}
\put(-25,50){(e)}
\vspace*{-.05cm}
\end{subfigure}
\begin{subfigure}[b]{0.32\textwidth}
\centering
\includegraphics[width=\textwidth]{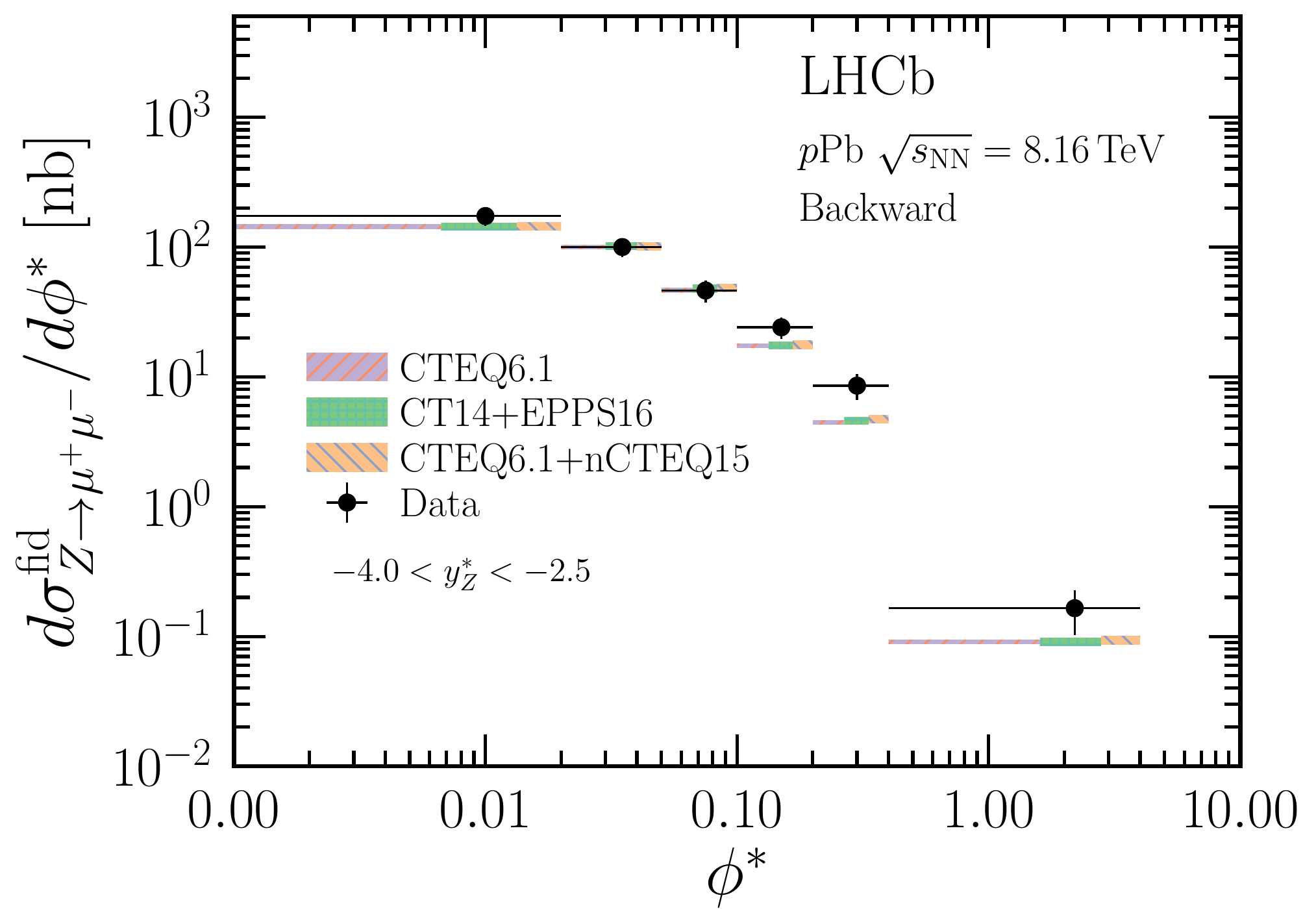}
\put(-25,50){(f)}
\end{subfigure}
\end{center}
\vspace*{-0.5cm}
\caption{
Differential cross-section as a function of (left column) \zrapstar, (middle column) \ZpT and (right column) \phistar,
together with the theoretical predictions,
where the top row is for forward collisions and the bottom row is for backward collisions.
}
\vspace*{-.5cm}
\label{fig:zmmxsec}
\end{figure}

\begin{figure}[htbp]
\begin{center}
\begin{subfigure}[b]{0.32\textwidth}
\centering
\includegraphics[width=\textwidth]{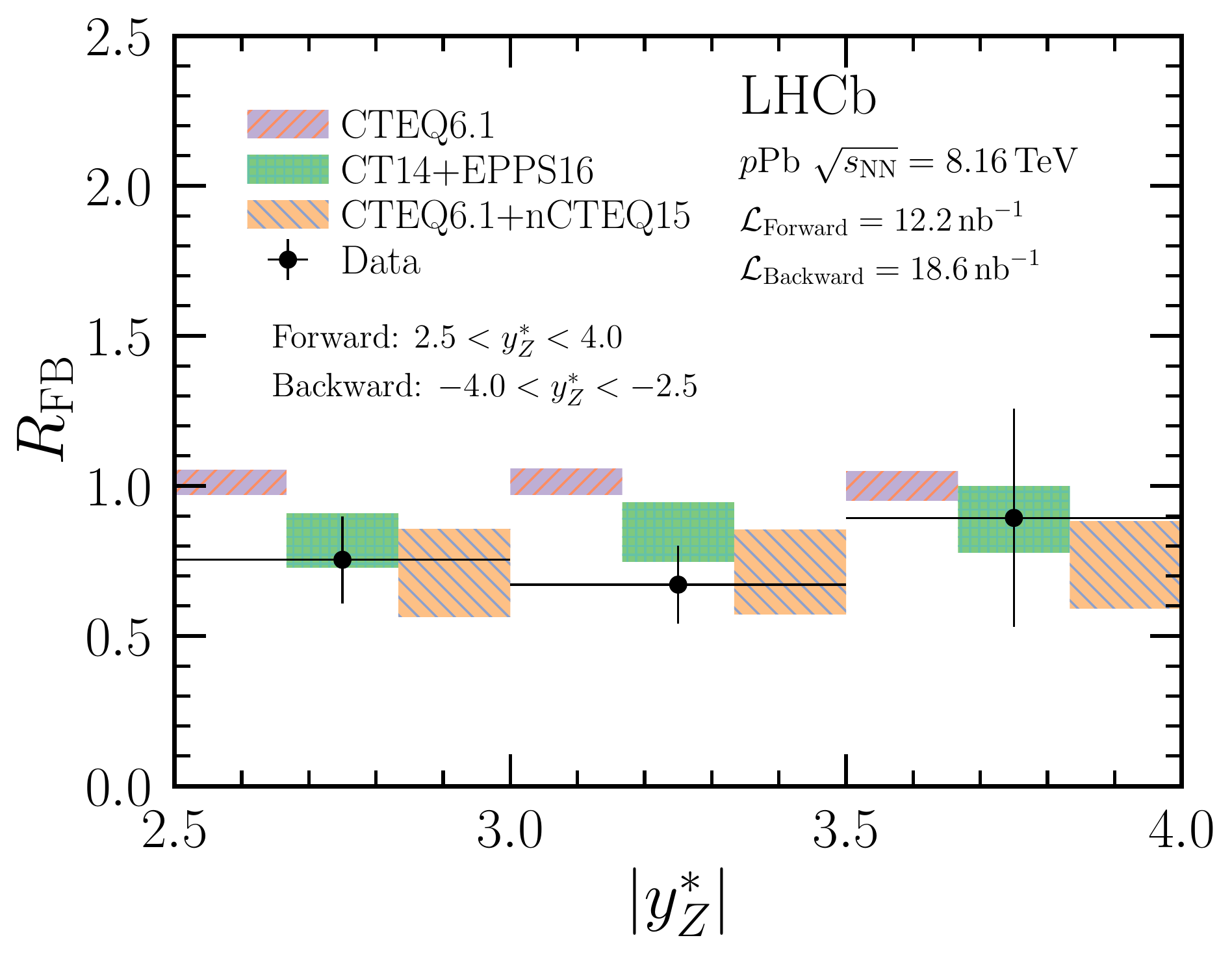}
\put(-40,50){(a)}
\end{subfigure}
\begin{subfigure}[b]{0.31\textwidth}
\centering
\includegraphics[width=\textwidth]{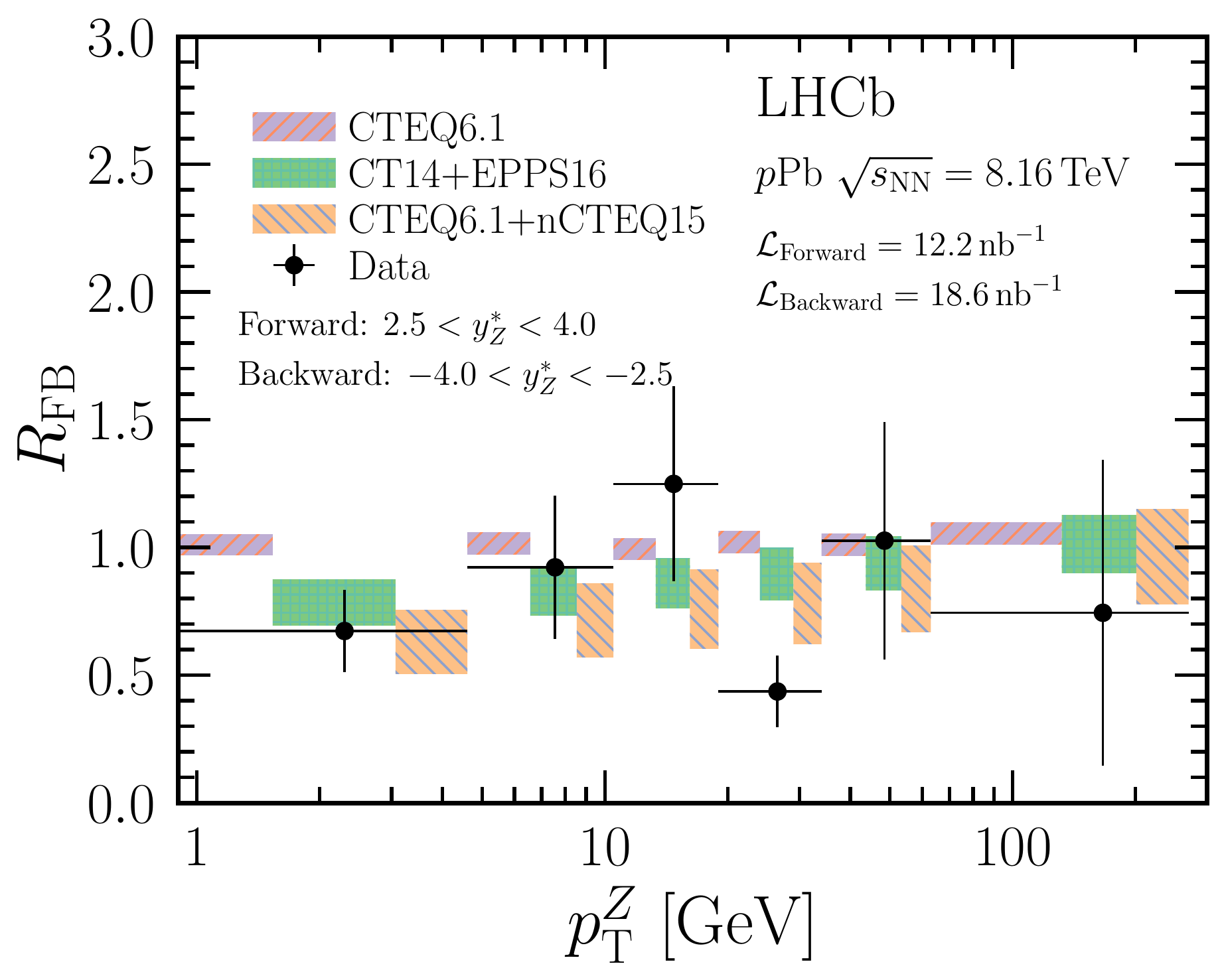}
\put(-40,52){(b)}
\vspace*{-.05cm}
\end{subfigure}
\begin{subfigure}[b]{0.32\textwidth}
\centering
\includegraphics[width=\textwidth]{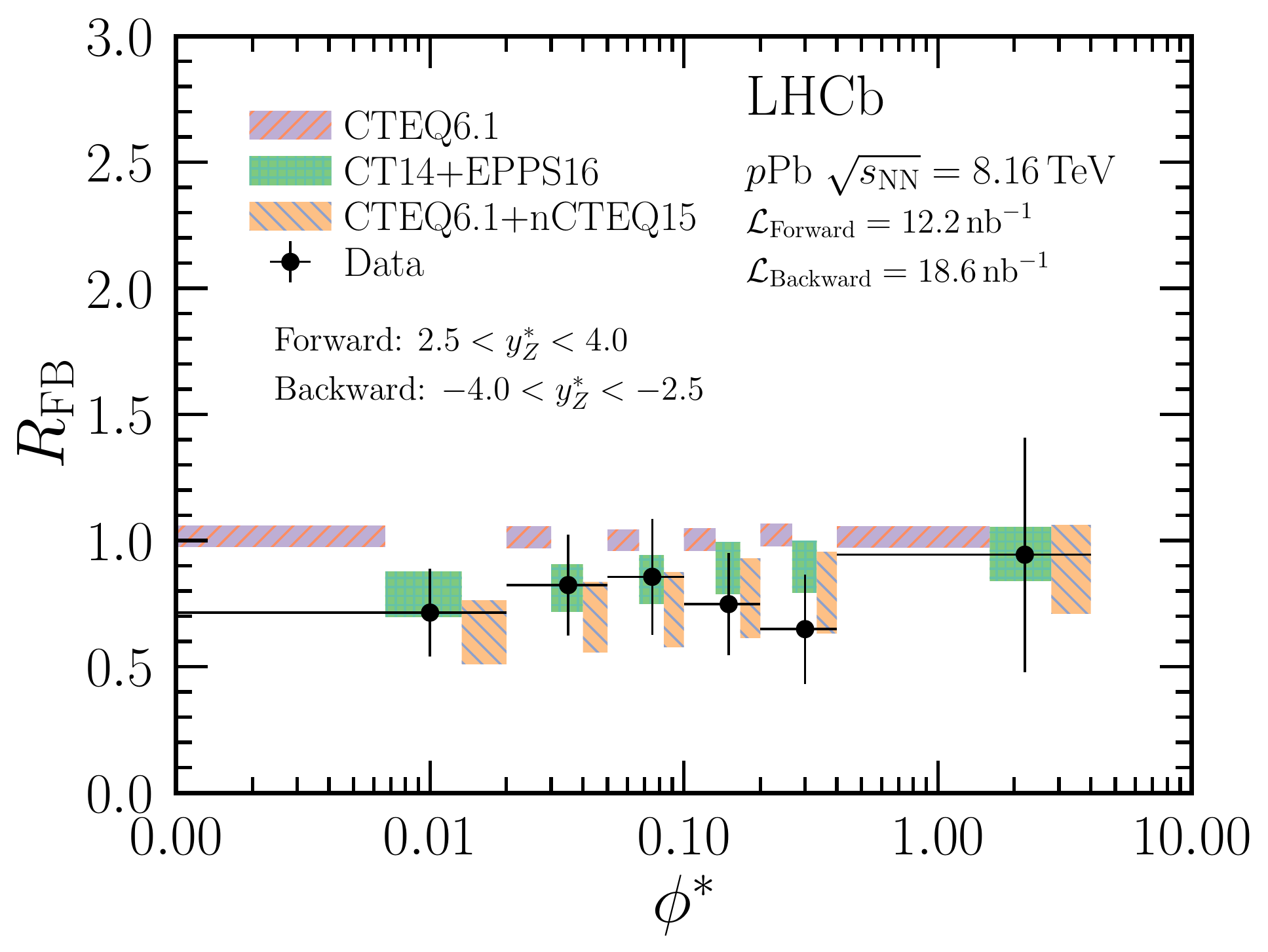}
\put(-40,50){(c)}
\end{subfigure}
\end{center}
\vspace*{-.5cm}
\caption{
Forward-backward ratio (\rfb) as a function of (left column) \zrapstar, (middle column) \ZpT and (right column) \phistar,
together with the theoretical predictions,
where the top row is for forward collisions and the bottom row is for backward collisions.
}
\vspace*{-.5cm}
\label{fig:zmmrfb}
\end{figure}

\begin{figure}[htbp]
\begin{center}
\begin{subfigure}[b]{0.32\textwidth}
\centering
\includegraphics[width=\textwidth]{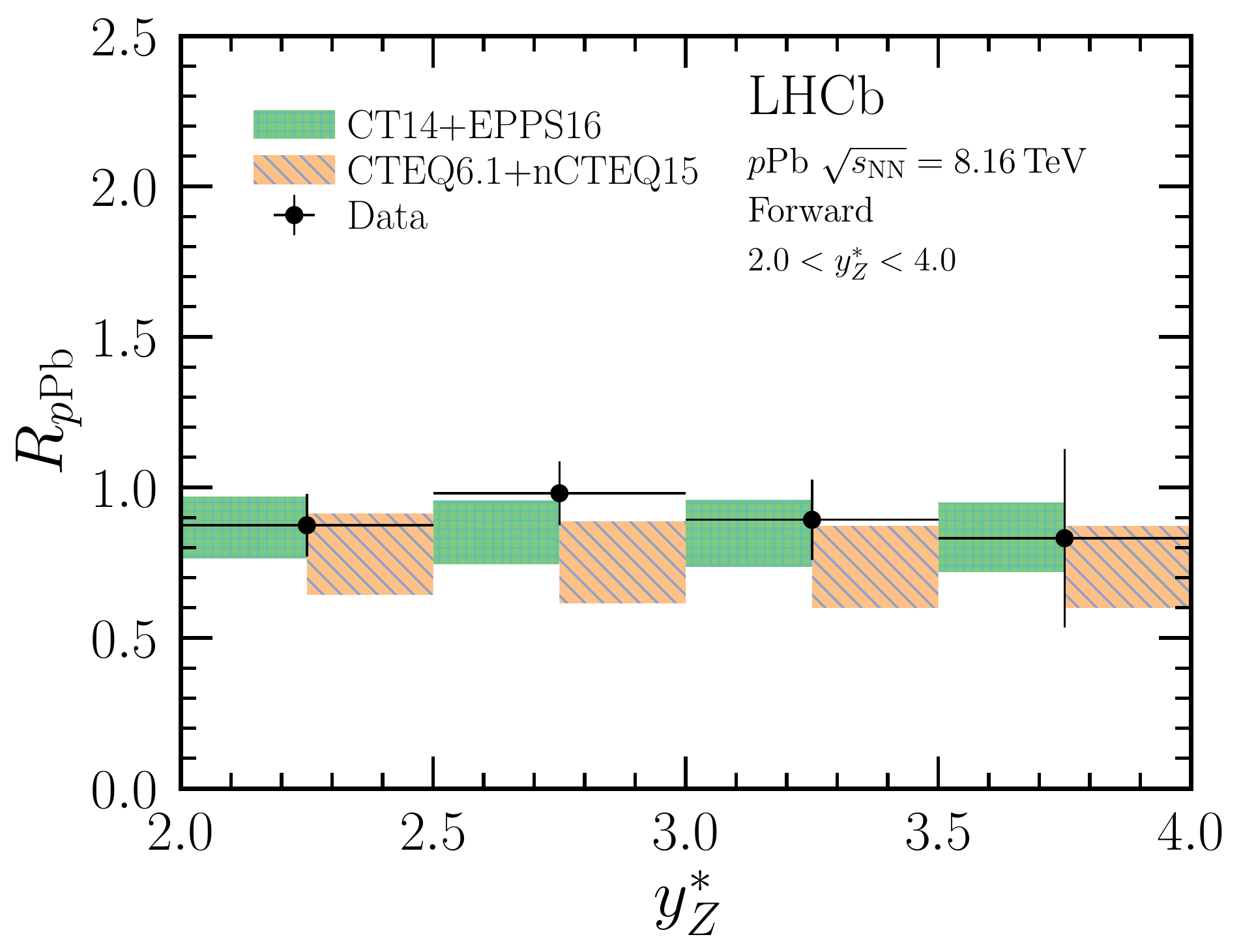}
\put(-30,20){(a)}
\end{subfigure}
\begin{subfigure}[b]{0.30\textwidth}
\centering
\includegraphics[width=\textwidth]{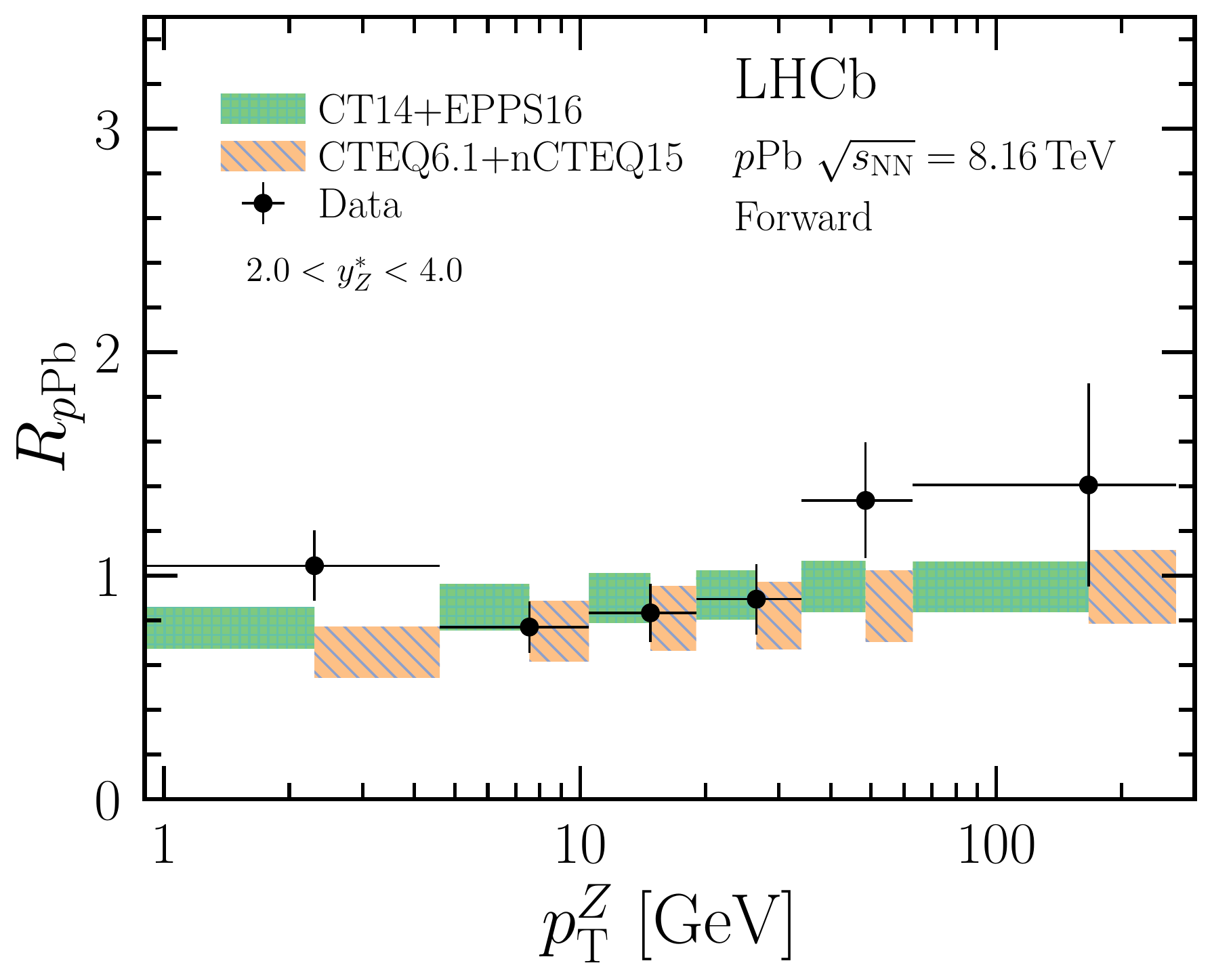}
\put(-30,25){(b)}
\vspace*{-.07cm}
\end{subfigure}
\begin{subfigure}[b]{0.32\textwidth}
\centering
\includegraphics[width=\textwidth]{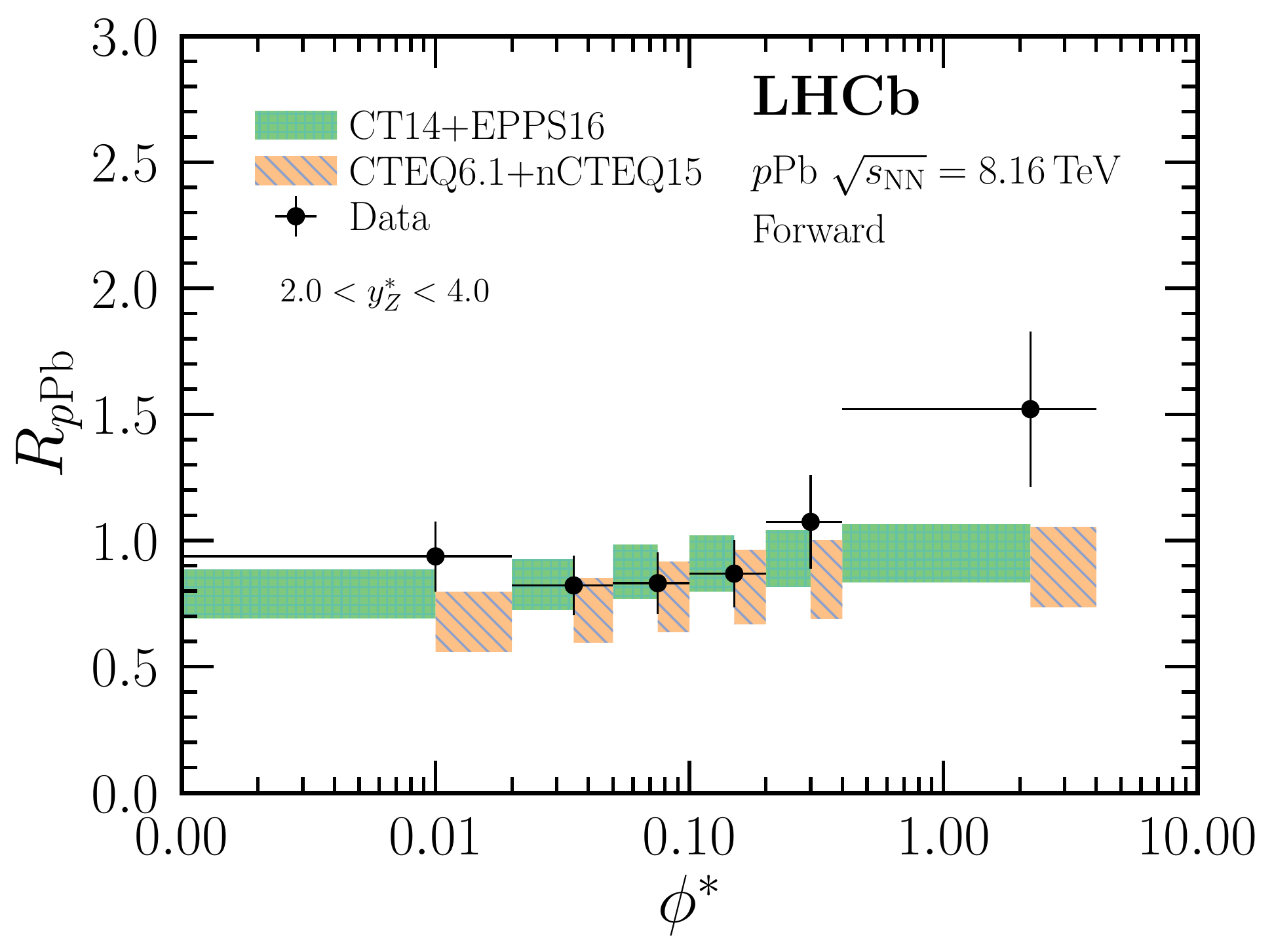}
\put(-30,20){(c)}
\end{subfigure}
\begin{subfigure}[b]{0.325\textwidth}
\centering
\includegraphics[width=\textwidth]{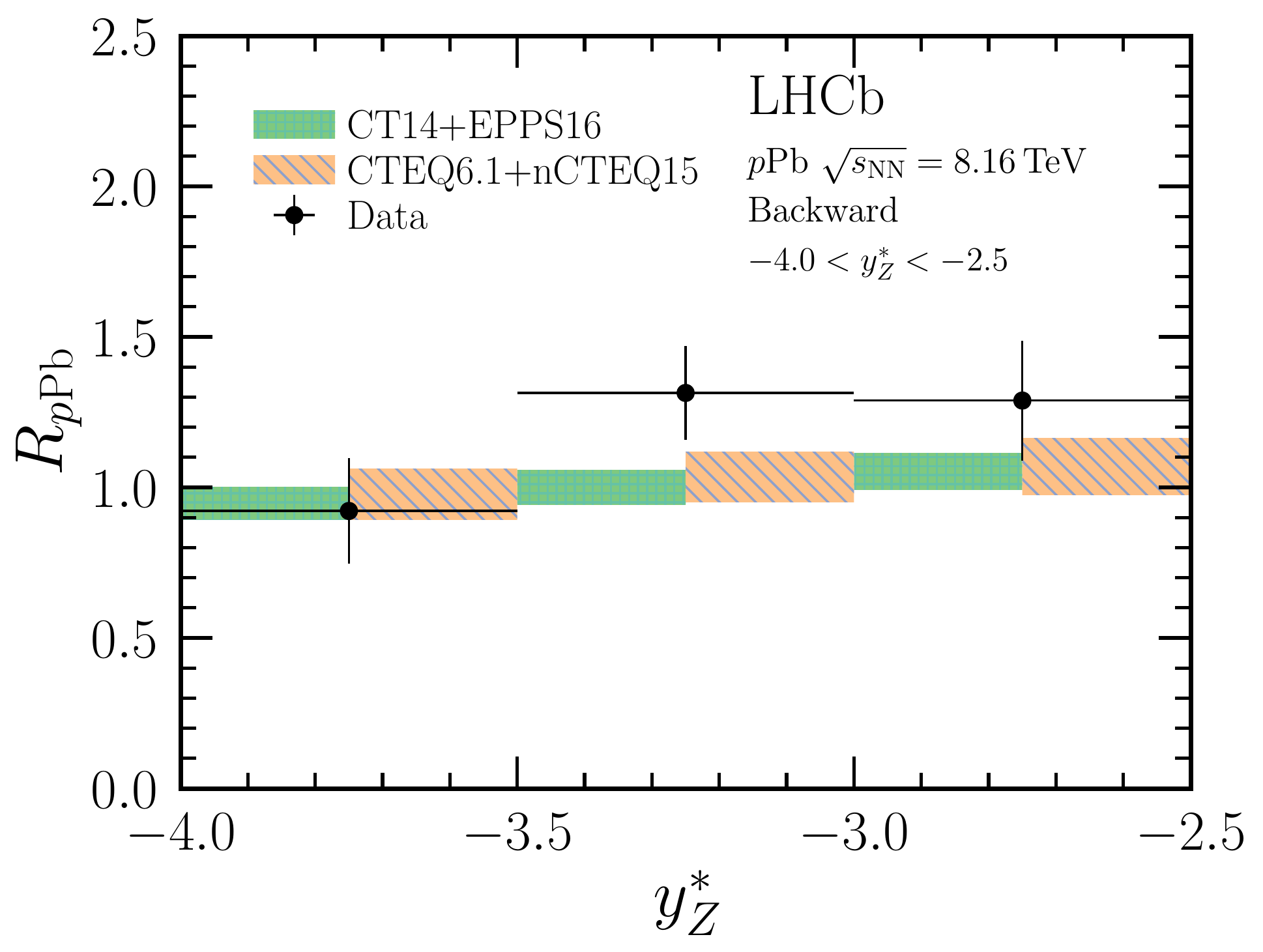}
\put(-30,20){(d)}
\end{subfigure}
\begin{subfigure}[b]{0.30\textwidth}
\centering
\includegraphics[width=\textwidth]{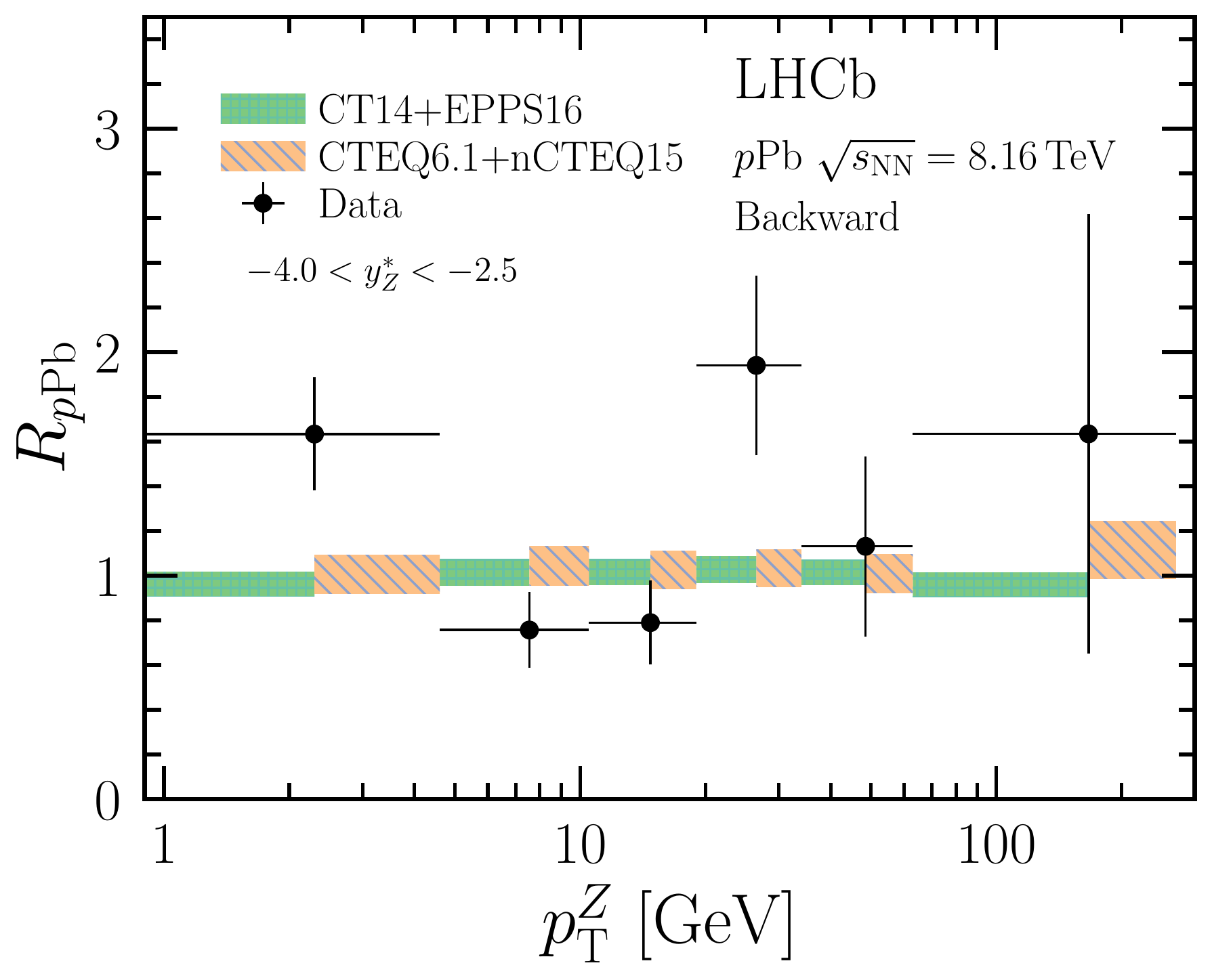}
\put(-30,25){(e)}
\vspace*{-.07cm}
\end{subfigure}
\begin{subfigure}[b]{0.32\textwidth}
\centering
\includegraphics[width=\textwidth]{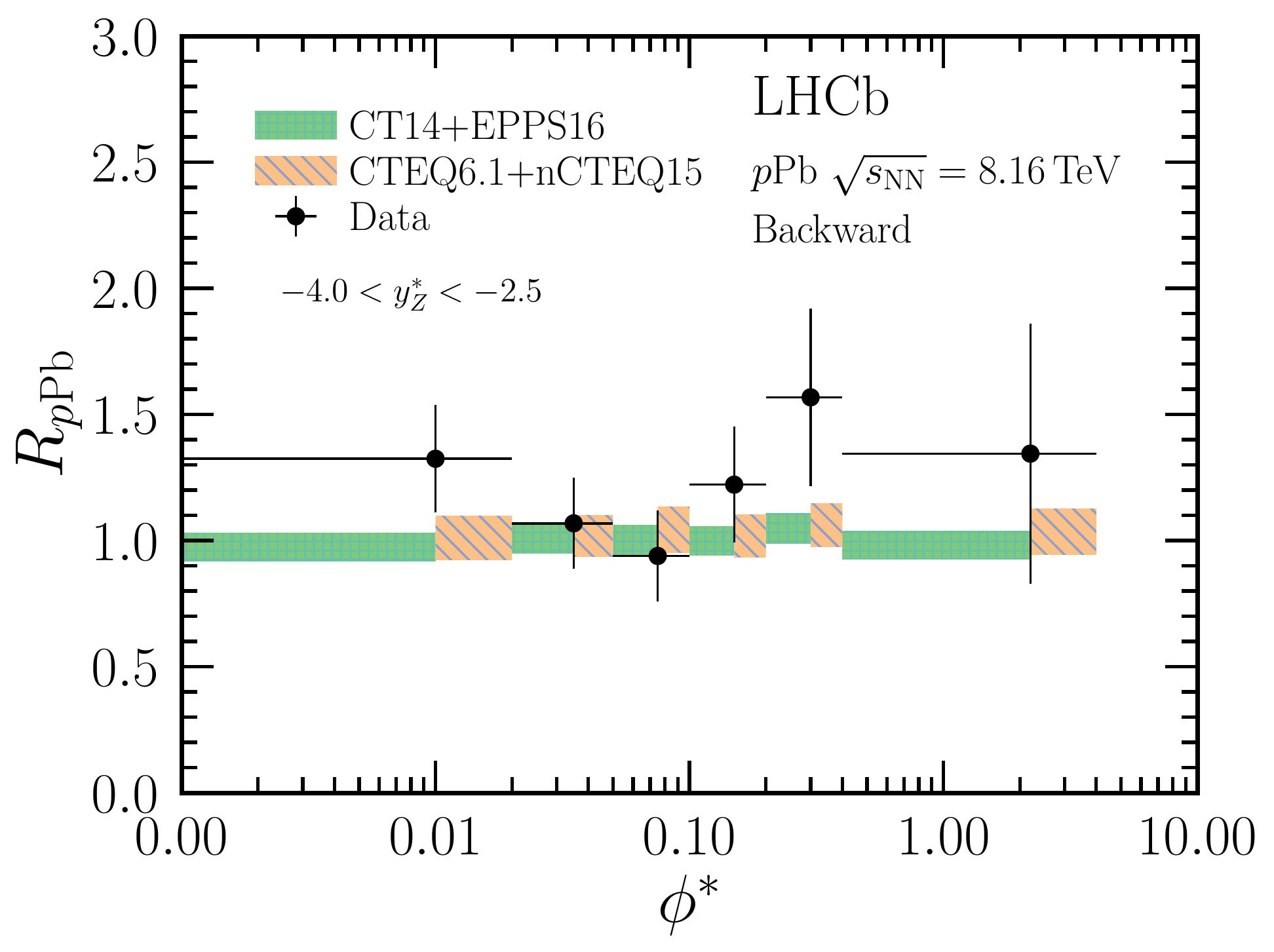}
\put(-30,20){(f)}
\end{subfigure}
\end{center}
\vspace*{-.5cm}
\caption{
Nuclear modification factors (\rpa) as a function of (left column) \zrapstar, (middle column) \ZpT and (right column) \phistar,
together with theoretical predictions,
where the top row is for forward collisions and the bottom row is for backward collisions.
}
\vspace*{-.5cm}
\label{fig:zmmrpa}
\end{figure}

The differential measurements of the production cross-sections, \rfb and \rpa are shown in Figs.~\ref{fig:zmmxsec}, \ref{fig:zmmrfb} and \ref{fig:zmmrpa}, respectively. 
In general, these results are compatible 
with nPDF predictions.
For the results of the cross-section and \rpa, larger uncertainties compared to the current nPDF predictions appear in the backward collisions for 
all three observables.
However, for forward rapidity
the large dataset gives a higher precision for certain intervals
compared to the nPDF predictions.
For the measured \rpa results, an overall suppression below unity as expected can be observed. However, given the current large uncertainties of the measurements, no conclusive statement can be made.

\section{Summary}

The exclusive coherent \jpsi and \psitwos
production and their cross-section ratio in UPC \PbPb collisions are newly measured using 2018 dataset collected by the LHCb detector.
This is the most precise coherent \jpsi production measurement at the moment, the first coherent \psitwos measurement, and the first ratio measurement between \psitwos and \jpsi productions, in forward rapidity region for UPC at LHC.
The differential cross-section of coherent \jpsi and \psitwos production in \PbPb UPC is also measured as a function of $\pt^*$ for the first time.
The measurements are compatible with various theoretical predictions, the high precision of the data is of great value for the fine-tuning of theoretical models. 
The \Z-boson production is measured at $\sqsnn=8.16\tev$ using \pPb collision dataset collected by the LHCb detector.
The differential cross-section, \rfb and \rpa as a function 
of \zrapstar, \ZpT, and \phistar are measured for the first time in the forward region at LHCb.
The new results are compatible with nCTEQ15 or EPPS16 nPDFs calculations.
Forward (small Bjorken-x) results show strong constraining power on the nPDFs.

\bibliographystyle{JHEP}
\bibliography{main}

\end{document}